\documentclass{jpsj3}
\usepackage{times}
\usepackage[T1]{fontenc}
\usepackage{textcomp}
\def\e{{\rm e}}

\def\ggs{\buildrel\textstyle > \over {\hbox{\raise0.2ex\hbox{$\sim$}}}}
\def\lls{\buildrel\textstyle < \over {\hbox{\raise0.2ex\hbox{$\sim$}}}}
\def\gsim{\,\lower0.75ex\hbox{$\ggs$}\,}
\def\lsim{\,\lower0.75ex\hbox{$\lls$}\,}

\def\tt{\tilde{t}}
\def\im{{\rm i}}

\def\e{{\rm e}}

\def\jo #1#2#3#4{#1 {\bf #2} (#3) #4}   %For J. Phys. Soc. Jpn.
\def\PRB{Phys.\ Rev.\ B}
\def\PRL{Phys.\ Rev.\ Lett.}

\def\JPSJ{J.\ Phys.\ Soc.\ Jpn.}

\def\PHE{Physica E}

\def\IJMPB{Int.\ J.\ Mod.\ Phys.\ B}

\def\EPJB{Eur.\ Phys.\ J.\ B}
\title{Anomalous Interaction Dependence in Magnetism of Graphene Nanoribbons with
Zigzag Edges}

\author{Karin \textsc{Furukawa},  
Hideo \textsc{Yoshioka}\thanks{E-mail address:
h-yoshi@cc.nara-wu.ac.jp}, and
Yoneko \textsc{Mochizuki}}  

\inst{Department of Physics, Nara Women's University, Nara 630-8506}

\recdate{\today}

\abst{Properties in magnetic ordered states 
of graphene nanoribbons with zigzag shaped edges are investigated by 
applying mean-field approximation to the Hubbard model with on-site repulsion $U$.  
We observe that magnetic moments and critical temperature 
show anomalous power-law dependences as a function of $U$; the actual
values of the power are determined by only the width of ribbons. 
Such singular behaviours are found to be due to localized nature of
the electronic states close to Fermi energy.    
}

\kword{graphene nanoribbon, zigzag shaped edges, edge states, magnetism}

\begin{document}
\maketitle

\section{Introduction} %% No sections necessary for express letters, letters and short notes

Graphene-based materials with nano-meter sizes have been attracting much
attention due to their possibilities as new potential devices in
application  
as well as novel stages for emergence of exotic phenomena in fundamental science. 
Especially, the graphene nanoribbon with zigzag shaped edges, 
which is abbreviated to zigzag GNR in the following, is known to
 have fascinating peculiar
properties as follows.\cite{KKobayashi,Fujita,Nakada,Wakabayashi,Miyamoto,Brey,Sasaki1} 
The zigzag GNR has a metallic band structure irrespective of the width
$N$ in the sense that the energy gap does not appear at the Fermi
energy in the absence of doping, $E=0$. 
However, unlike usual metals, 
the asymptotic form of the energy dispersion near $E=0$ 
is written as $E \propto \pm |k-\pi/a|^N$,\cite{Wakabayashi}  
and the Fermi velocity vanishes for $N \geq 2$ 
where $a$ and $N$ express the lattice
spacing and the width of the ribbon (see
Fig.\ref{fig:model}), respectively.     
Such characteristic properties are due to the fact that 
the one-particle states near $E=0$ are well localized around zigzag
edges, i.e., the states close to $E=0$ are so called edge states.   
The zigzag edges and the localized states around them have been observed
by scanning tunneling microscopy and spectroscopy.\cite{Kobayashi1,Niimi1,Kobayashi2,Niimi2,science-edge-1,science-edge-2}

Magnetic properties of the zigzag GNR have been investigated by applying
mean-field approximation to the Hubbard model 
with on-site repulsion $U$ 
\cite{Fujita,Wakabayashi2,Rossier,Sasaki2,Jung} and by the first
principles calculation.\cite{Kusakabe,Lee,Son,Pisani,Yazyev}
It has been found that the large spontaneous magnetic moments appear at
the zigzag edges, which is originated from the edge states. 
The magnetic moments align ferromagnetically at each edge but with
the opposite direction between the edges.   
In addition, the magnetic order appears under the infinitesimal on-site repulsion;  
the conclusion is different from the case for graphene sheets  
where the finite amount of $U$ is necessary for emergence of 
the antiferromagnetic state. 
The spin excitations\cite{Wakabayashi2,Yazyev} and the interedge
superexchange interaction\cite{Jung} have been calculated based on the
magnetic structure introduced above.  
The treatments beyond the mean-field approximation, in which quantum
fluctuation is fully taken into account, have been carried
out.\cite{Hikihara,Yoshioka,Hajj} 
The ground state is found to be Mott insulator with charge gap, and 
the field theoretical approach demonstrates that   
the Heisenberg model on the zigzag GNR expressing the spin excitation 
belongs to the same universality class as spin 1/2 square ladders\cite{Yoshioka}
(gapped for even number legs, gapless for odd number of legs).
The result is confirmed by numerical calculation.\cite{Hajj}   

In ordered states seen in electron systems, 
usually, the order parameter shows exponential
dependence as a function of the coupling constant; 
this fact is due to the finite density of states (DOS) at the Fermi energy. 
On the other hand, the DOS of the zigzag GNR show divergence at 
Fermi energy. 
Therefore, in the magnetic ordered state of the zigzag GNR, 
unusual dependence of the spontaneous magnetic moments as a
function of $U$ is expected.   
In the present work, 
we study properties of the magnetic ordered states in the zigzag GNR by
applying the mean-field approximation to the Hubbard model.  
We focus on interaction dependence of spontaneous magnetic moments 
and critical temperature. 
It is found  that these quantities show anomalous power-law dependences and the
actual values of the power are determined by only the width $N$ of the ribbons.   
We clarify that the unusual band structure close to Fermi energy
 gives rise to such singular behaviours. 

\section{Model}

\begin{figure}[htb]
\begin{center}
\includegraphics[width=10.0truecm]{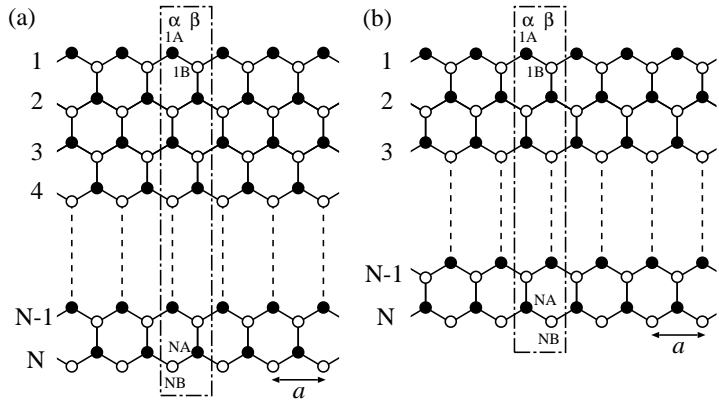}
\end{center}
\caption{
Schematic illustration of zigzag GNRs consisting of $N$ legs 
with  $N =$ even (a) and $N =$ odd (b). 
Here, the rectangle written by the dash dotted line shows the unit cell
 and $a$ is the lattice spacing. 
The filled (open) circles express the A (B) sublattices. 
The two types of slices in the unit cell are denoted as $\alpha$ and
 $\beta$. 
}
\label{fig:model}
\end{figure}

The graphene nanoribbons with zigzag shaped edges we study
are illustrated in a schematic way in Fig. \ref{fig:model}. 
%--------------- 
%--------------
Figs. \ref{fig:model} (a) and (b) show the zigzag GNRs with the width $N$ being even and
odd, respectively. 
In the respective figures, the rectangle written by the dash dotted line and $a$ express 
the unit cell and the lattice spacing, respectively. 
The filled (open) circles indicate the A (B) sublattices. 
In the unit cell, there are two types of carbon slices; those are
denoted as $\alpha$ and $\beta$.    
 
The Hamiltonian we consider reads 
$\mathcal{H} = \mathcal{H}_{\rm k} + \mathcal{H}_{\rm int}$, where
%-----------------
\begin{align}
 \mathcal{H}_{\rm k} =& -t \sum_{I,\sigma} \sum_{i=1}^{N} 
\left\{
c^\dagger_{iB,\sigma} (I) c_{iA,\sigma} (I) 
+ c^\dagger_{iA,\sigma} (I) c_{iB,\sigma} (I+(-1)^i) + \mathrm{h.c.}     
\right\} \nonumber \\
& -t \sum_{I,\sigma} \sum_{i=1}^{N-1} 
\left\{ 
c^\dagger_{iB,\sigma} (I) c_{(i+1)A,\sigma} (I) + \mathrm{h.c.}     
\right\}, \\
%--------------
 \mathcal{H}_{\rm int} =& \frac{U}{2} \sum_{I,\sigma} \sum_{i=1}^N 
\left\{
n_{iA,\sigma}(I) n_{iA,-\sigma}(I) + n_{iB,\sigma}(I) n_{iB,-\sigma}(I) 
\right\}, 
\end{align}
%--------------
where $t$ is the hopping between the nearest neighbor carbon atoms. 
Here $c^\dagger_{iX,\sigma} (I)$ is
the creation operator of an  electron with the spin $\sigma = \pm$
($\sigma = +/-$ expresses $\uparrow/\downarrow$ spin state)
at the $iX$ site in the $I$-th cell, and 
$n_{iX,\sigma} (I) = c^\dagger_{iX,\sigma} (I) c_{iX,\sigma} (I)$
($i=1, \cdots, N$ and $X=A,B$ unless explicitly noted in the following). 
The mean-field approximation is applied to $\mathcal{H}_{\rm int}$ as
%----------------
\begin{align}
\mathcal{H}_{\rm int} \to \mathcal{H}_{\rm int}^{\rm MF} =&
 \frac{U}{2} \sum_{I,\sigma} \sum_{i=1}^N 
\big\{
 ( n_{iA} - \sigma m_{iA}) c^\dagger_{iA,\sigma}(I) c_{iA,\sigma}(I) \nonumber \\
& + ( n_{iB} - \sigma m_{iB}) c^\dagger_{iB,\sigma}(I) c_{iB,\sigma}(I)
\big\},
\end{align}
%-----------------
where $n_{iX} = \sum_\sigma \langle n_{iX,\sigma} (I)
\rangle$ and $m_{iX} = \sum_\sigma \sigma \langle n_{iX,\sigma} (I)
\rangle$ are the charge and spin order parameters 
with $\langle \cdots \rangle$ being the thermal average. 
We should note that the magnetic solution 
as well as the paramagnetic one in the neutral system  
has $n_{iA} = n_{iB} = 1$ due to the particle-hole symmetry. 
Therefore, in the following, we neglect the terms including the charge
order parameters for simplicity since those renormalize the chemical potential.     
As a result, the Hamiltonian under the mean-field approximation is written as follows,
%---------------
\begin{align}
 \mathcal{H}^{\rm MF} = \sum_{k,\sigma} \Psi^\dagger (k,\sigma)
\left\{ h_{\rm k}(k) + h_{\rm int}^{\rm MF} (\sigma) \right\}
 \Psi(k,\sigma), 
\end{align} 
%-------------
where $\Psi^\dagger(k,\sigma) = (c^\dagger_{1A,\sigma}(k), c^\dagger_{1B,\sigma}(k), \cdots ,
c^\dagger_{NA,\sigma}(k), c^\dagger_{NB,\sigma}(k))$.
Here $h_{\rm k}(k)$ and
$h_{\rm int}^{\rm MF} (\sigma)$ are the $2N \times 2N$ matrices, 
%--------------
\begin{align}
h_{\rm k} (k) 
&= -t \times \left(
\begin{array}{ccccccccc}
 0 & \gamma_k & 0 & 0 & \cdots & 0 & 0 & 0 & 0 \\
 \gamma_k & 0 & 1 & 0 & \cdots & 0 & 0 & 0 & 0 \\
 0 & 1 & 0 & \gamma_k & \cdots & 0 & 0 & 0 & 0  \\
 0 &  0 & \gamma_k & 0 & \cdots & 0 & 0 & 0 & 0  \\
 \vdots & \vdots & \vdots & \vdots & \ddots & \vdots & \vdots & \vdots & \vdots \\
 0 & 0 & 0 & 0 & \cdots & 0 & \gamma_k & 0 & 0 \\
 0 & 0 & 0 & 0 & \cdots & \gamma_k & 0 & 1 & 0 \\
 0 & 0 & 0 & 0 & \cdots & 0 & 1 & 0 & \gamma_k \\
 0 & 0 & 0 & 0 & \cdots & 0 & 0 & \gamma_k  & 0 \\
\end{array}
\right), 
\label{eqn:hk}
\\
h_{\rm int}^{\rm MF} (\sigma) 
&= - \sigma \frac{U}{2} \times  {\rm diag} 
\left(
 m_{1A}, m_{1B}, \cdots , m_{NA}, m_{NB}
\right),
\end{align}
%--------------
where $\gamma_k = 2 \cos (ka/2)$. 
Here, the Fourier transformations
%--------------
\begin{align}
 c_{iA,\sigma} (I) &= \frac{1}{\sqrt{N_L}} \sum_k \e^{\im k (x_I+(-1)^i
 a/4)} c_{iA,\sigma}(k), 
\label{eqn:FA}
\\
 c_{iB,\sigma} (I) &= \frac{1}{\sqrt{N_L}} \sum_k \e^{\im k (x_I-(-1)^i
 a/4)} c_{iB,\sigma}(k),
\label{eqn:FB}
\end{align}
%---------------
are introduced with $N_L$ being the total number of the unit cell in the
system and $x_I = Ia$.
We note that $h_{\rm k} (k)$ becomes the real and symmetric matrix 
owing to the choice of the Fourier transformation, eqs. \eqref{eqn:FA}
and \eqref{eqn:FB}.   
The $2N$ order parameters $m_{iA}$ and $m_{iB}$ are determined
self-consistently by
%-------------- 
\begin{align}
 m_{iA} &= \frac{1}{N_L} \sum_{k,\sigma} \sum_{j=1}^{2N} \sigma 
\left\{ \left[ \vec{v}^{(j)}_\sigma (k) \right]_{2i-1} \right\}^2
 f(E^{(j)}_\sigma (k)), \\
 m_{iB} &= \frac{1}{N_L} \sum_{k,\sigma} \sum_{j=1}^{2N} \sigma 
\left\{ \left[ \vec{v}^{(j)}_\sigma (k) \right]_{2i} \right\}^2
 f(E^{(j)}_\sigma (k)), 
\end{align}
%----------------
where $f(E)$ is the Fermi function defined by $f(E) = 1/ \{ \exp(E/T) + 1 \}$ with $T$ being the temperature.  
Here $E^{(j)}_\sigma (k)$ is an eigenvalue of the matrix $h_{\rm k}(k)
+ h_{\rm int}^{\rm MF}(\sigma)$ and 
the corresponding eigenvector is expressed by
$\vec{v}^{(j)}_\sigma (k)$, the $l$-th ($l=1,2,\cdots,2N$) element of
which is a real number and written as 
$\left[ \vec{v}^{(j)}_\sigma (k) \right]_{l}$. 
The energy per an atom $\epsilon (U)$ is given by
\begin{align}
 \epsilon (U) = \frac{1}{2 N_L N} \sum_{k,\sigma} \sum_{j=1}^{2N}
 E^{(j)}_\sigma (k) f(E^{(j)}_\sigma (k)) + \frac{U}{8N} \sum_{i=1}^{N}
\left( m_{iA}^2 + m_{iB}^2 \right).
\label{eqn:E}
\end{align} 

\section{Results and Discussions}
\begin{figure}[htb]
\begin{center}
%\includegraphics[width=7.0truecm]{F-AF-N3.eps} \\
%\hspace{-0.2cm}\includegraphics[width=7.1truecm]{F-AF-N3-ENERGY.eps} 
\includegraphics[width=7.1truecm]{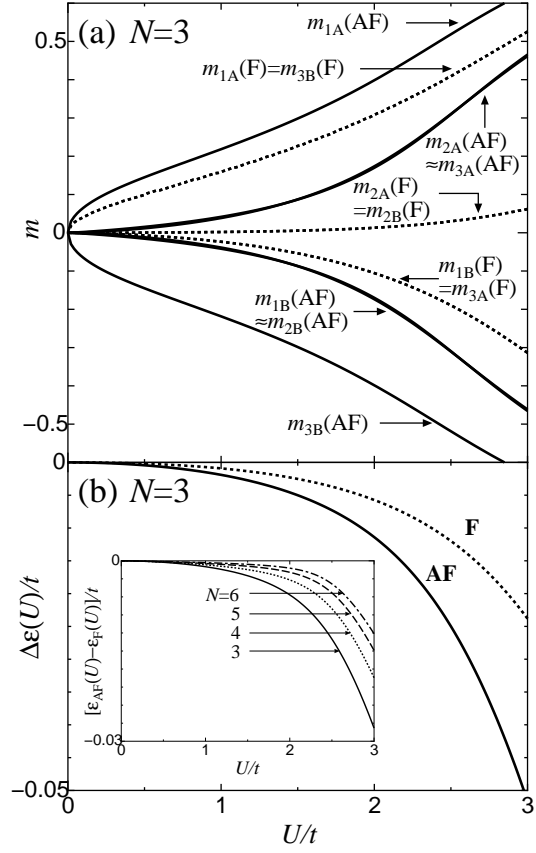} 
\end{center}
\caption{
Self-consistent solutions of the AF state and the F state 
for the $N=3$ zigzag GNR; 
magnetic moments at each atom in the unit cell, $m_{iA}$ and $m_{iB}$, (a) 
and energy difference  
$\Delta \epsilon (U) = \epsilon (U) - \epsilon (0)$ (b) 
where $\epsilon (U)$ is defined in eq. (\ref{eqn:E}). 
In each figure, the solid and dotted curves express the quantities 
in the AF state and those for the F state, respectively.  
The inset in (b) shows the energy difference between the AF
 state and the F state, $\epsilon_\mathrm{AF} (U) -
 \epsilon_\mathrm{F} (U)$, 
for several choices of $N$.  
}
\label{fig:N3}
\end{figure}

We can obtain the two kinds of self-consistent solutions;  
one expresses the antiferromagnetic (AF) state which satisfies 
$m_{iA} = - m_{(N-i+1)B}$ and 
the other is the ferromagnetic (F) state with $m_{iA} = m_{(N-i+1)B}$.\cite{Lee,Pisani,Jung} 
The magnetic moment at each site 
and the energy difference $\Delta \epsilon (U) = \epsilon (U) - \epsilon (0)$ 
in the both states are shown in Figs. \ref{fig:N3} (a) and (b), respectively, 
for the $N=3$ system at the absolute zero temperature. 
%--------------- 
%--------------
In each figure, the quantities in the AF state and those in the F state
are expressed by the solid and dotted curves, respectively.  
Though difference between $m_\mathrm{2A}$ and $m_\mathrm{3A}$ 
($m_\mathrm{1B}$ and $m_\mathrm{2B}$) in the AF state is not clearly 
seen in Fig. \ref{fig:N3} (a), 
there exists the significant difference between the both quantities.   
As we expected, the AF state is more stable than the F state.\cite{Pisani} 
Then, we concentrate on the AF state unless noted. 
Note that the energy difference per an atom between 
the AF state and the F state 
$\epsilon_\mathrm{AF} (U) - \epsilon_\mathrm{F} (U)$
becomes smaller
with increasing the width $N$. 
This fact seems to indicate that 
difference in the energy is mainly originated from that in the local
spin structure, 
for example, the spin configuration at the bond in the center of the ribbon, 
i.e., $n$A-$n$B bond for $N=2n-1$ and $n$B-($n+1$)A for $N=2n$. 
Qualitative discrepancies between the $N=$odd
zigzag GNR and the $N=$even one have been observed 
in transport properties\cite{Wakabayashi-Aoki,Akhmerov,Li,Cresti,Nakabayashi,Ranis,Mochizuki-1,Mochizuki-2} and persistent currents.\cite{Yoshioka-Higashibata}  
Such discrepancies are not found in the magnetic ordered states obtained
by the mean-field approximation.   

\begin{figure}[htb]
\begin{center}
\includegraphics[width=7.0truecm]{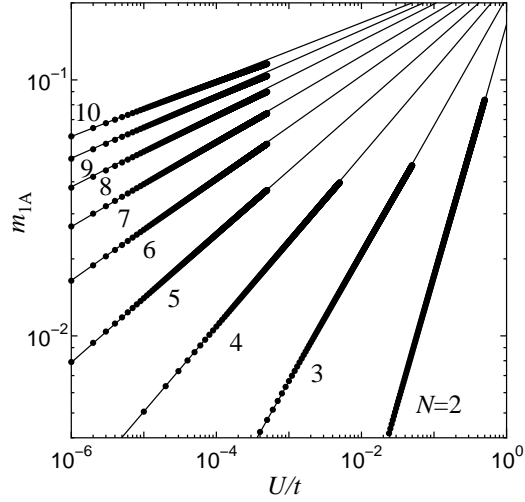}
\end{center}
\caption{Magnetic moment at the 1A site, $m_\mathrm{1A}$, of the AF state 
as a function of $U/t$ for the several choices of $N$. 
The solid lines express the fitting by $m_\mathrm{1A} \propto (U/t)^{1/(N-1)}$.} 
\label{fig:edge}
\end{figure}

We investigate in detail the magnetic moment at the zigzag edges, 
which is far largest than the others, and then considered to dominate
the magnetic properties as far as $U/t \ll 1$.     
The magnetic moment at the 1A site $m_\mathrm{1A}$ of the AF state  
is shown as a function of $U/t$ close to $U/t = 0$ 
in Fig. \ref{fig:edge}
for $N= 2, 3, \cdots, 10$.          
%--------------- 
%--------------- 
We can see that the magnetic moment $m_\mathrm{1A}$ behaves as 
$m_\mathrm{1A} \propto (U/t)^{1/(N-1)}$, i.e., 
the quantity shows power-law dependences and   
the actual value of the power is determined by the width $N$. 
The result seems to be anomalous 
because the order parameters of the usual ordered states realized in the
electronic systems are known to
show exponential dependences as a function of the coupling constant
for weak coupling limit. 
Actually, in the case of $N=1$ which is nothing but the usual
one-dimensional system, $m_{1A} = -m_{1B} \simeq 16 (t/U) \exp(-2 \pi
t/U)$ as long as $U/t \ll 1$.    

%--------------
\begin{figure}[htb]
\begin{center}
\includegraphics[width=7.0truecm]{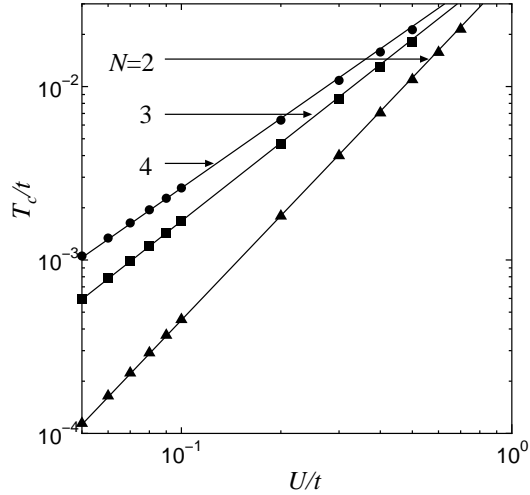}
\end{center}
\caption{
Critical temperature $T_c$ of the AF state as a function of $U/t$ 
for the several choices of $N$. 
The solid lines express the fitting by $T_c/t \propto (U/t)^{N/(N-1)}$.}
\label{fig:TC}
\end{figure}
%---------------
Next we investigate the critical temperature $T_c$ of the AF state,
which is also known to show exponential dependence as a function of  the coupling
constant in the usual ordered state.       
Fig. \ref{fig:TC} shows the critical temperature as a function of $U/t$
close to $U/t = 0$. 
Here, the power-law dependence is also seen in the critical temperature as
$T_c \propto (U/t)^{N/(N-1)}$. 
The actual value of the power in $T_c$ is different from that 
in $m_\mathrm{1A}$. 
However, it can be well understood by the simple dimension analysis, 
$T_c \sim U \times m_\mathrm{1A}$,  
with considering that the AF transition is dominated by the edges
states.  

Here, we explore the critical temperature from divergence 
of the magnetic susceptibility. 
The instability toward the AF state and that toward the F state will be investigated.   
In the presence of the external magnetic field $H_{iX}$, 
which does not depend on an index of the unit cell $I$ but does depend
on the location in the unit cell, 
the linear response theory results in the magnetic moments,
$m_{iA}$ and $m_{iB}$ as
%----------------- 
\begin{align}
m_{iA} &= \sum_{i'=1}^N \left\{ \chi_{iA,i'A}^0 H_{i'A}^{\rm eff} +
 \chi_{iA,(N - i' +1)B}^0 H_{(N - i'+1)B}^{\rm eff}  \right\}, 
\label{eqn:RRTA}
\\
m_{(N-i+1)B} &= \sum_{i'=1}^N \big\{ \chi_{(N - i +1)B,i'A}^0
 H_{i'A}^{\rm eff} +
 \chi_{(N-i+1)B,(N - i'+1)B}^0 H_{(N - i'+1)B}^{\rm eff}  \big\}, 
\label{eqn:RRTB}
\end{align}
%-----------------
where $H_{iX}^{\rm eff} = H_{iX} + U m_{iX}/2$ is the effective magnetic
field at the $iX$ cite. 
The quantities $\chi_{iX,i'X'}^0 = \lim_{q \to 0} \lim_{\omega_n \to 0}
\chi_{iX,i'X'}^0 (q, \im \omega_n)$ ($X,X' = A,B$)
is the susceptibility of the non-interacting system:
%------------------ 
\begin{align}
\chi_{iA,i'A}^0 (q, \im \omega_n) =& - \frac{2}{N_L} \sum_k \sum_{j,j'=1}^{2N} 
\e^{\im q \left\{ (-1)^i - (-1)^{i'} \right\} a/4}
\frac{ f(E^{(j)}_0(k)) - f(E^{(j')}_0(k+q)) }{ \im \omega_n + E^{(j)}_0(k) - E^{(j')}_0(k+q) }
 \nonumber \\
&\times 
\left[ \vec v^{(j)}_0(k)\right]_{2i'-1}
\left[ \vec v^{(j)}_0(k)\right]_{2i-1}
\left[ \vec v^{(j')}_0(k+q)\right]_{2i-1}
\left[ \vec v^{(j')}_0(k+q)\right]_{2i'-1}, \\
\chi_{iA,i'B}^0 (q, \im \omega_n) =& - \frac{2}{N_L} \sum_k \sum_{j,j'=1}^{2N} 
\e^{\im q \left\{ (-1)^i + (-1)^{i'} \right\} a/4}
\frac{ f(E^{(j)}_0(k)) - f(E^{(j')}_0(k+q)) }{ \im \omega_n + E^{(j)}_0(k) - E^{(j')}_0(k+q) }
 \nonumber \\
&\times 
\left[ \vec v^{(j)}_0(k)\right]_{2i'}
\left[ \vec v^{(j)}_0(k)\right]_{2i-1}
\left[ \vec v^{(j')}_0(k+q)\right]_{2i-1}
\left[ \vec v^{(j')}_0(k+q)\right]_{2i'}, \\
\chi_{iB,i'A}^0 (q, \im \omega_n) =& - \frac{2}{N_L} \sum_k \sum_{j,j'=1}^{2N} 
\e^{\im q \left\{ - (-1)^i - (-1)^{i'} \right\} a/4}
\frac{ f(E^{(j)}_0(k)) - f(E^{(j')}_0(k+q)) }{ \im \omega_n + E^{(j)}_0(k) - E^{(j')}_0(k+q) }
 \nonumber \\
&\times 
\left[ \vec v^{(j)}_0(k)\right]_{2i'-1}
\left[ \vec v^{(j)}_0(k)\right]_{2i}
\left[ \vec v^{(j')}_0(k+q)\right]_{2i}
\left[ \vec v^{(j')}_0(k+q)\right]_{2i'-1}, \\
\chi_{iB,i'B}^0 (q, \im \omega_n) =& - \frac{2}{N_L} \sum_k \sum_{j,j'=1}^{2N} 
\e^{\im q \left\{ - (-1)^i + (-1)^{i'} \right\} a/4}
\frac{ f(E^{(j)}_0(k)) - f(E^{(j')}_0(k+q)) }{ \im \omega_n + E^{(j)}_0(k) - E^{(j')}_0(k+q) }
 \nonumber \\
&\times 
\left[ \vec v^{(j)}_0(k)\right]_{2i'}
\left[ \vec v^{(j)}_0(k)\right]_{2i}
\left[ \vec v^{(j')}_0(k+q)\right]_{2i}
\left[ \vec v^{(j')}_0(k+q)\right]_{2i'}, 
\end{align}
%-------------------------
with $E^{(j)}_0(k)$ ($j=1,\cdots,2N$) being 
the eigenvalue of $h_{\rm k}(k)$ 
and $\vec v^{(j)}_0(k)$ being the eigenvector corresponding to it. 
Namely, the element of $\vec v^{(j)}_0(k)$ expresses 
the amplitude of the $j$-th eigenfunction 
in the transverse direction in the non-interacting case.   
 
In the F state, 
the magnetic moments satisfy the configuration $m_{iA} = m_{(N-i+1)B}$, 
whereas $m_{iA} = - m_{(N-i+1)B}$ is realized in the AF state.  
In order to lead to such a magnetic structure,  
the effective external field should satisfy 
$H_{iA}^{\rm eff} = H_{(N-i+1)B}^{\rm eff}$ 
( $H_{iA}^{\rm eff} = - H_{(N-i+1)B}^{\rm eff}$ )
for the F (AF) state.       
Based on the above consideration, eqs. (\ref{eqn:RRTA}) or
(\ref{eqn:RRTB})
are rewritten as follows, 
%-------------- 
\begin{align}
 \vec m_{A} &= \tilde \chi_{F/AF} \vec H_A, \\
 \tilde \chi_{F/AF} &= \left( \tilde 1 - \frac{U}{2} \tilde
 \chi^0_{F/AF}\right)^{-1} \tilde \chi_{F/AF}^0,
\end{align}
%--------------
where 
$\vec m_{A} = (m_{1A},m_{2A},\cdots,m_{NA})^{\rm t}$, 
$\vec H_{A} = (H_{1A},H_{2A},\cdots,H_{NA})^{\rm t}$ 
and $\tilde 1$ is the $N \times N$ unit matrix. 
Here, $\tilde \chi^0_{F}$ and $\tilde \chi^0_{AF}$ are the $N \times N$
matrices
whose element is defined as
\begin{align}
 \left[ \tilde \chi^0_{F/AF} \right]_{ii'} &= \chi^0_{iA,i'A} \pm
 \chi^0_{iA,(N-i'+1)B} \nonumber \\
&= - \frac{2}{N_L} \sum_k \sum_{j,j'=1}^{2N} 
\lim_{q \to 0} 
\frac{ f(E^{(j)}_0(k)) - f(E^{(j')}_0(k+q)) }{ E^{(j)}_0(k) - E^{(j')}_0(k+q) }
 \nonumber \\
&\times 
\left[ \vec v^{(j)}_0(k)\right]_{2i-1}
\left[ \vec v^{(j')}_0(k)\right]_{2i-1}  \nonumber \\
&\times 
\left(
\left[ \vec v^{(j)}_0(k)\right]_{2i'-1} 
\left[ \vec v^{(j')}_0(k)\right]_{2i'-1} 
\pm 
\left[ \vec v^{(j)}_0(k)\right]_{2(N-i'+1)} 
\left[ \vec v^{(j')}_0(k)\right]_{2(N-i'+1)}  
\right), 
\end{align}  
where the upper and lower sign correspond to 
$ \left[ \tilde \chi^0_{F} \right]_{ii'}$ and
$ \left[ \tilde \chi^0_{AF} \right]_{ii'}$, respectively. 
The transition temperature is determined by $\det ( \tilde 1 - (U/2) \tilde
\chi^0_{F/AF}) = 0 $. 

\begin{figure}[htb]
\begin{center}
\includegraphics[width=7.0truecm]{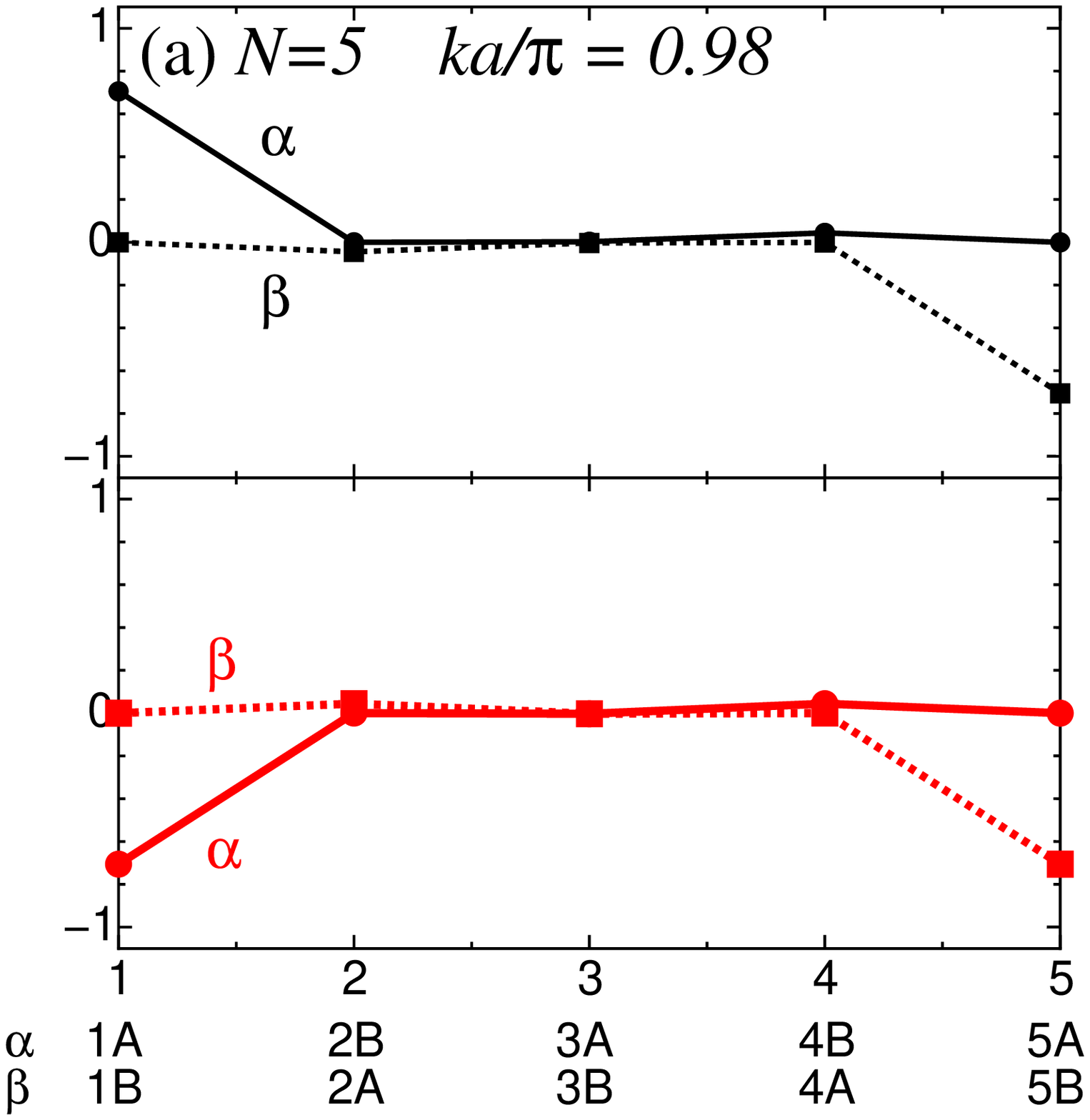} \\
\includegraphics[width=7.0truecm]{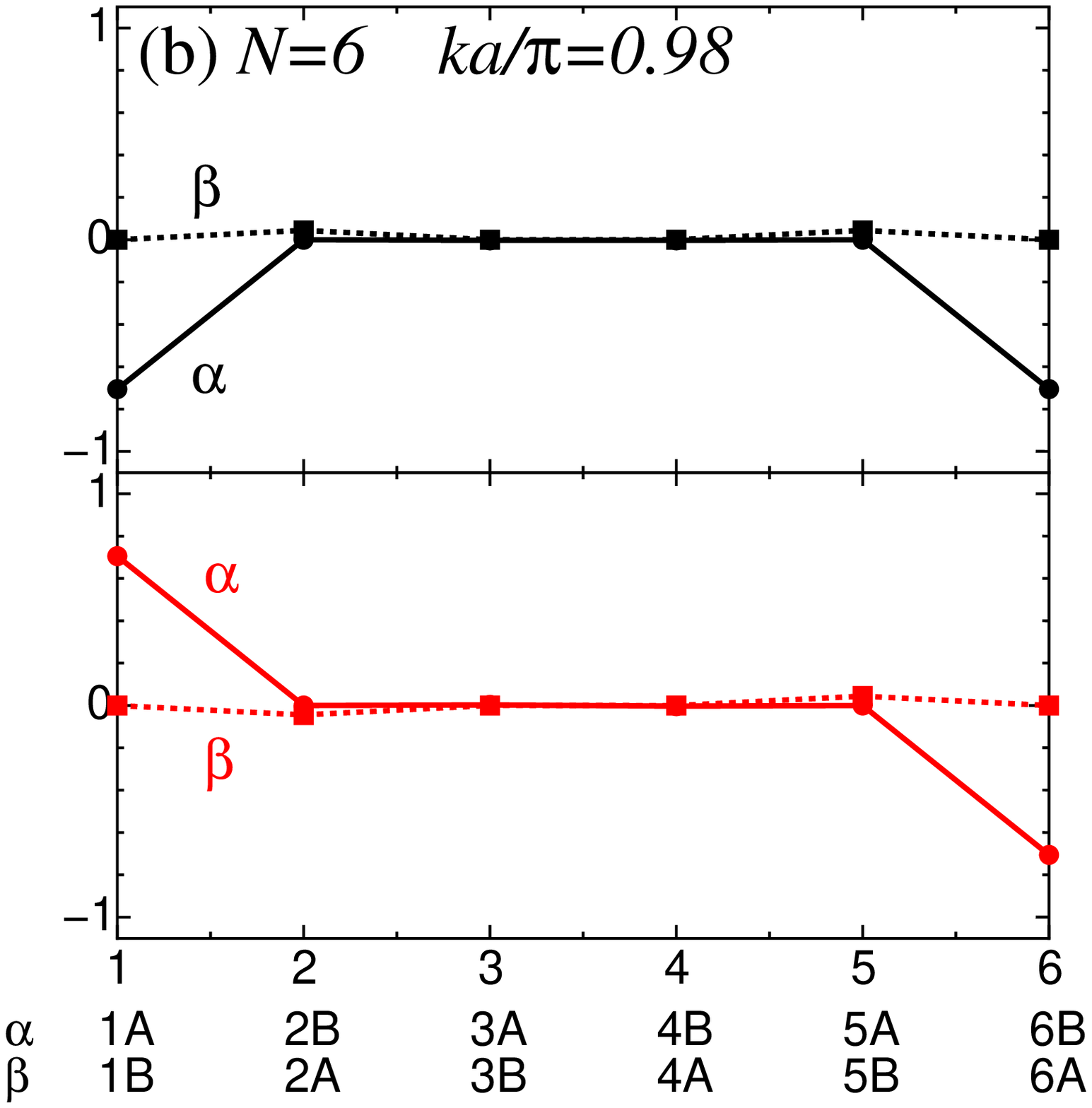} 
\end{center}
\caption{
(Color online) Amplitudes of the eigenfunctions in the transverse direction close to $E/t
 = 0$ for $N=5$ (a) and $N=6$ (b) where 
$ka/\pi = 0.98$ is used. 
In the respective figure, the upper and lower graph express the wavefunction
 of the conduction band and of the valence band, respectively. 
The slices $\alpha$ and $\beta$ in the unit cell are defined in Fig. \ref{fig:model}.    
}
\label{fig:wf}
\end{figure}

We focus on the transition temperature for weak repulsion, in which case 
it is numerically obtained that the transition temperature is proportional
to $U^{N/(N-1)}$. 
In this case, 
we take account of only the two kinds of eigenstates whose eigenvalues are  
around $E=0$, 
i.e., $E = E_0^{(N+1)} = - E_0^{(N)} \simeq A_N |k - \pi/a|^N$ with 
$A_N$ being a positive constant.   
Eigenfunctions of such states are known to be well localized
around the zigzag edges, which are explicitly shown in
Fig. \ref{fig:wf}.  
%----------------
%----------------
At the sufficiently low temperature, 
owing to the localized nature of the eigenfunctions, 
the matrix elements of $\tilde \chi^0_{F/AF}$ can be
neglected except $[ \tilde \chi^0_{F/AF} ]_{11}$, 
which are calculated as follows,
%------------ 
\begin{align}
 [ \tilde \chi^0_{F} ]_{11} &\simeq - \frac{2}{N_L} \sum_k \left. \frac{\partial
 f}{\partial \epsilon}\right|  _{\epsilon = E_0^{(N+1)}(k)}  = 2
 \int_0^\infty d \epsilon \left\{ - \frac{\partial f}{\partial \epsilon}
 \right\} D_N (\epsilon), 
\label{eqn:chiF11}
\\
 [ \tilde \chi^0_{AF} ]_{11} &\simeq - \frac{2}{N_L} \sum_k
 \left. \frac{ f(\epsilon) - f(- \epsilon) }{2
 \epsilon}\right|_{\epsilon = E_0^{(N+1)}(k)}
=  
2 \int_0^\infty d \epsilon \left\{ \frac{f(-\epsilon) - f(\epsilon)}{2 \epsilon}
 \right\} D_N (\epsilon).
\label{eqn:chiAF11}  
\end{align} 
%-------------
Here, $D_N (\epsilon)$, which is the density of states close to 
$\epsilon = 0$, is obtained as $D_N (\epsilon) \simeq C_N/|\epsilon|^{1-1/N}$~~\cite{Wakabayashi} from 
the asymptotic behaviour of the energy dispersion, 
$E_0^{(N+1)} = - E_0^{(N)} = A_N | k - \pi/a |^N$.   
In deriving eqs. (\ref{eqn:chiF11}) and (\ref{eqn:chiAF11}), 
the amplitude at the zigzag edges are approximated as   
$\left[ \vec{v}_0^{(j)} (k) \right]_1 = \pm 1/\sqrt{2}$ and 
$\left[ \vec{v}_0^{(j)} (k) \right]_{2N} = \pm 1/\sqrt{2}$ ($j=N,N+1$), 
and the others are discarded. 
In addition, the sign of two kinds of amplitudes are     
 assigned according to the result in Fig. \ref{fig:wf}, 
e.g., in the odd $N$ case, 
$\left[ \vec{v}_0^{(N+1)} (k) \right]_1 = 1/\sqrt{2}$,  
$\left[ \vec{v}_0^{(N+1)} (k) \right]_{2N} = - 1/\sqrt{2}$,  
$\left[ \vec{v}_0^{(N)} (k) \right]_1 = -1/\sqrt{2}$,  and 
$\left[ \vec{v}_0^{(N)} (k) \right]_{2N} = - 1/\sqrt{2}$ 
from Fig. \ref{fig:wf} (a).     
The transition temperature $T_{c,\mathrm{F/AF}}$, which is determined by 
$1- (U/2)[\tilde \chi^0_{F/AF}]_{11} = 0$, is obtained as follows;
\begin{align}
 T_{c,{\rm F}} &= \left( C_N X_{{\rm F},N} U \right)^{N/(N-1)}, \\
 T_{c,{\rm AF}} &= \left( C_N X_{{\rm AF},N} U \right)^{N/(N-1)}, 
\end{align} 
where $X_{{\rm F/AF},N}$ is an constant depending on the width $N$,
\begin{align}
 X_{{\rm F},N} &= \int_0^\infty dx \frac{1}{\e^x + 1} \frac{1}{\e^{-x} +
 1} \frac{1}{x^{1-1/N}}, \\
 X_{{\rm AF},N} &= \int_0^\infty dx \frac{1}{\e^x + 1} \frac{1}{\e^{-x}+1}
 \frac{1}{x^{1-1/N}} \frac{\sinh x}{x}.
\end{align} 
Thus the transition temperatures of the both magnetic states are
proportional to $U^{N/(N-1)}$; 
the result for $T_{c,{\rm AF}}$ is identical with that
obtained by the numerical calculation.
Note that $T_{c,{\rm AF}} > T_{c,{\rm F}}$ because of $X_{{\rm AF},N} >
X_{{\rm F},N}$, which corresponds to the fact that the antiferromagnetic state is
more stable than the ferromagnetic state. 
 
\begin{figure}[htb]
\begin{center}
\includegraphics[width=7.0truecm]{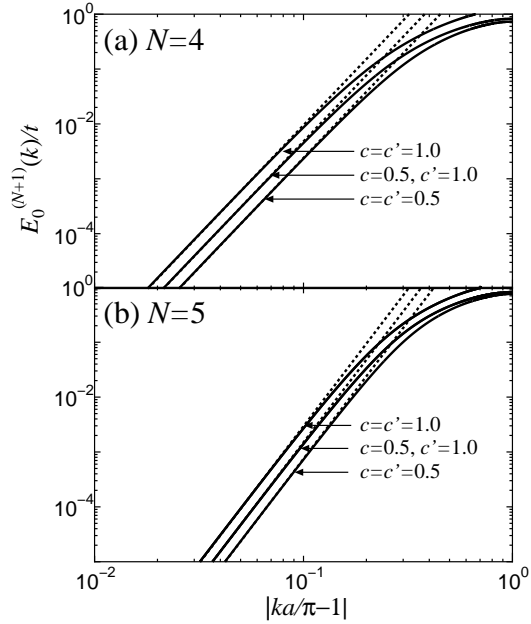} 
\end{center}
\caption{
Energy dispersion $E_0^{(N+1)} (k)$ $(= - E_0^{(N)}(k))$ close to $k=\pi/a$
 for several choices of $c$ and $c'$ in the case of $N=4$ (a) and $N=5$ (b). 
Here, the hopping at the 1st leg and at the $N$-th leg 
( see Fig. \ref{fig:model} ) are modified as
 $c \times t$ and $c' \times t$, respectively. 
In each figure, the dotted line express the fitting by using
 $E_0^{(N+1)}(k) \propto |k - \pi/a|^N$.  
}
\label{fig:asymband}
\end{figure}

Finally, we discuss the case where the hopping integral at the 1st leg 
and that of the $N$-th
 one are modified as $c \times t$ and $c' \times t$, respectively. 
In this case, the matrix elements of $h_{\rm k}(k)$ in
eq. (\ref{eqn:hk}) are changed as 
$[h_{\rm k}(k)]_{1,2} = [h_{\rm k}(k)]_{2,1} = - ct \gamma_k$ and 
$[h_{\rm k}(k)]_{2N-1,2N} = [h_{\rm k}(k)]_{2N,2N-1} = - c't \gamma_k$. 
Even upon such a change, 
the power of the energy dispersion chose to $E=0$ is not changed; 
i.e., the energy dispersion is given  by 
$E_0^{(N+1)} = - E_0^{(N)} \sim A_N (c,c') |k-\pi/a|^N$ as is shown in  
Fig. \ref{fig:asymband}.%\cite{Mochizuki} 
%----------------
%----------------
%----------------
\begin{figure}[htb]
\begin{center}
\includegraphics[width=7.0truecm]{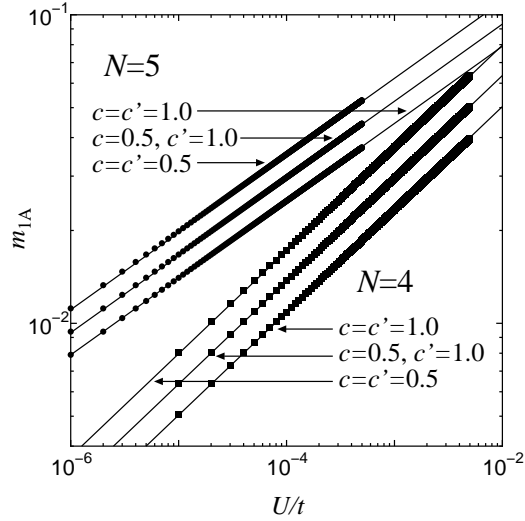} 
\end{center}
\caption{ Spontaneous magnetic moment at the 1A site 
(upper zigzag edge in Fig. \ref{fig:model}) $m_{1A}$
of the AF state with $N=4$ and $N=5$ at $T=0$ 
as a function of $U/t$ for several choices of $c$ and $c'$.  
The solid lines express the fitting by $m_\mathrm{1A} \propto
 U^{1/(N-1)}$.   
}
\label{fig:asym_m}
\end{figure}
%---------------- 
The power-law dependence of the magnetic moment at the edge 
$m_\mathrm{1A} \propto U^{1/(N-1)}$ should be
observed even in the present situation
if the anomalous power-law dependence of the magnetic moments and of 
the critical
temperature discussed above 
are originated from the dispersion relation close to $E/t=0$. 
In Fig. \ref{fig:asym_m}, we show the magnetic moments at the 1A site of
the AF state at $T=0$ for the $N=4$ and $N=5$ system 
as a function of $U/t$ for several choices of $c$ and $c'$. 
Here, the power-law dependence $m_\mathrm{1A} \propto U^{1/(N-1)}$ 
is observed even if the hopping integrals at the edges are modified. 
The result is the strong evidence that the anomalous power-law dependence
found in the present work is originated from the power-law dependence of the energy
dispersion close to $E/t=0$. 
We note that the magnetic moments at the edges does not satisfy 
the simple relation $m_{1A} = - m_{NB}$ and are obtained as $m_{1A} > |m_{NB}|$  
in the asymmetric case with $c < c'$ 
though the total magnetic moment in the unit cell vanishes. 
Even in such a case, as long as $U/t \ll 1$, 
$|m_{NB}|$ is also proportional to $U^{1/(N-1)}$ as well as $m_{1A}$.

\section{Summary}
In the present work, 
we applied the mean-field approximation to the Hubbard model on the
zigzag GNR and studied properties of the magnetic ordered states. 
The spontaneous magnetic moments at the zigzag edges 
and the transition temperature of the AF state were investigated in
detail as a function of on-site repulsion $U$ for $U/t \ll 1$.  

We can obtain the two kinds of the ordered states; one is the
AF state satisfying $m_{iA} = - m_{(N-i+1)B}$ and 
the other is the F state with $m_{iA} = m_{(N-i+1)B}$. 
The AF state is more stable than the F state 
though the energy difference per one carbon atom between the two
magnetic ordered states becomes smaller with increasing the width $N$. 
Therefore, the AF state was investigated in detail. 
Due to existence of the states localized around the zigzag edges close
to Fermi energy, the magnetic moments at the zigzag edges are far 
bigger than the others for $U/t \ll 1$, and show characteristic $U/t$ 
dependence as $m_{1A} \propto (U/t)^{1/(N-1)}$. 
Also, the transition temperature $T_{c, \mathrm{AF}}$ shows 
the power-law dependence $T_{c, \mathrm{AF}} \propto (U/t)^{N/(N-1)}$, 
which is analytically demonstrated from the divergence of the corresponding
susceptibility. 
Discrepancy in the power of $m_{1A}$ and $T_{c, \mathrm{AF}}$ 
can be well understood by considering that the AF transition is
dominated by the edge states and by assuming the simple dimension analysis, 
$T_{c, \mathrm{AF}} \propto U \times m_{1A}$.  
Therefore, we can conclude that 
the both anomalous $U/t$ dependences are originated from the power-law
divergence of the DOS close to Fermi energy. 
Actually, the power-law dependence of the magnetic moment at the edges
are observed if the hopping integrals at the zigzag edges are modified, 
where the power of the DOS is unchanged.

\section*{Acknowledgment}
%The authors would like to thank Y. Mochizuki for valuable discussion. 
This work was supported by Nara Women's University Intramural Grant for Project Research.


\begin{thebibliography}{99} 
%% The number "99" means that this list has more than nine items.
%---------------
\bibitem{KKobayashi}
K. Kobayashi:
\jo{\PRB}{48}{1993}{1757}.
%----------------
\bibitem{Fujita}
M. Fujita, K. Wakabayashi, K. Nakada, and K. Kusakabe: 
\jo{\JPSJ}{65}{1996}{1920}.
%---------------
\bibitem{Nakada}
K. Nakada, M. Fujita, G. Dresselhaus, and M. S. Dresselhaus: 
\jo{\PRB}{54}{1996}{17954}.
%---------------
\bibitem{Wakabayashi}
K. Wakabayashi, M. Fujita, H. Ajiki, and M. Sigrist: 
\jo{\PRB}{59}{1999}{8271}.
%---------------
\bibitem{Miyamoto}
Y. Miyamoto, K. Nakada, and M. Fujita:
\jo{\PRB}{59}{1999}{9858}.
%---------------
\bibitem{Brey}
L. Brey and H.A. Fertig
\jo{\PRB}{73}{2006}{235411}.
%--------------
\bibitem{Sasaki1}
K. Sasaki, S. Murakami, and R. Saito:
\jo{\JPSJ}{75}{2006}{074713}.
%--------------


%---------------
\bibitem{Kobayashi1}
Y. Kobayashi, K. Fukui, T. Enoki, K. Kusakabe, and Y. Kaburagi:
\jo{\PRB}{71}{2005}{193406}.
%---------------
\bibitem{Niimi1}
Y. Niimi, T. Matsui, H. Kambara, K. Tagami, M. Tsukada, and H. Fukuyama:
\jo{Appl. Surf. Sci.}{241}{2005}{43}. 
%--------------
\bibitem{Niimi2}
Y. Niimi, T. Matsui, H. Kambara, K. Tagami, M. Tsukada, and H. Fukuyama:
\jo{\PRB}{73}{2006}{085421}.
%--------------
\bibitem{Kobayashi2}
Y. Kobayashi, K. Fukui, T. Enoki, and K. Kusakabe:
\jo{\PRB}{73}{2006}{125415}.
%-----------------
\bibitem{science-edge-1}
X. Jia, M. Hofmann, V. Meunier, B.G. Sumpter, J. Campos-Delgado, 
J.M. Romo-Herrera, H. Son, Y.-P. Hsieh, A. Reina, J. Kong, 
M. Terrones, and M.S. Dresselhaus: 
\jo{Science}{323}{2009}{1701}. 
%----------------
\bibitem{science-edge-2}
\k{C}. \"O. Girit, J. C. Meyer, R. Erni, M.D. Rossel, C. Kisielowskii,
	L. Yang, C.-H. Park. M.F. Crommie, M.L. Cohen, S.G. Louie and
	A. Zettl:  
\jo{Science}{323}{2009}{1705}. 

%------------- 
\bibitem{Wakabayashi2}
K. Wakabayashi, M. Sigrist, and M. Fujita:
\jo{\JPSJ}{67}{1998}{2089}.
%--------------
\bibitem{Sasaki2}
K. Sasaki and R. Saito:
\jo{\JPSJ}{77}{2008}{054703}.
%--------------
\bibitem{Rossier}
%J. Fern$\acute{\mathrm{a}}$ndez-Rossier: \jo{\PRB}{77}{2008}{075430}.
J. Fern{\'a}ndez-Rossier: \jo{\PRB}{77}{2008}{075430}.
%-------------
\bibitem{Jung}
J. Jung, T. Pereg-Barnea, and A.H. MacDonald:
\jo{\PRL}{102}{2009}{227205}.
 
%\bibitem{Kohn}
%W. Kohn: 
%\jo{\PR}{133}{1964}{A171}.
%--------------

\bibitem{Kusakabe}
K. Kusakabe and M. Maruyama:
\jo{\PRB}{67}{2003}{092406}.
%----------------
\bibitem{Lee}
H. Lee, Y.-W. Son, N. Park, S. Han, and J. Yu:
\jo{\PRB}{72}{2005}{174431}.
%----------------
\bibitem{Son}
Y.-W. Son, M.L. Cohen, and S.G. Louie:
\jo{\PRL}{97}{2006}{216803}.
%--------------
\bibitem{Pisani}
L. Pisani, J.A. Chan, B. Montanari, and N.M. Harrison:
\jo{\PRB}{75}{2007}{064418}.
%--------------
\bibitem{Yazyev}
Oleg V. Yazyev and M.I. Katsnelson:
\jo{\PRL}{100}{2008}{047209}.
%-------------
\bibitem{Yoshioka}
H. Yoshioka: 
\jo{\JPSJ}{72}{2003}{2145}.
%------------
\bibitem{Hikihara}
T. Hikihara, X. Hu, H.-H. Lin, and C.-Y. Mou:
\jo{\PRB}{68}{2003}{035432}.
%------------
\bibitem{Hajj}
M. Al Hajj, F. Alet, S. Capponi, M.B. Lepetit, J.-P. Malrieu, and
	S. Todo:
\jo{\EPJB}{51}{2006}{517}.
%--------------
\bibitem{Wakabayashi-Aoki}
K. Wakabayashi and T. Aoki:
\jo{\IJMPB}{16}{2002}{4897}.
%-------------
\bibitem{Akhmerov}
A. R. Akhmerov, J. H. Bardarson, A. Rycerz, and C. W. J. Beenakker: 
\jo{\PRB}{77}{2008}{205416}. 
%-------------
\bibitem{Li}
%Zuanyi Li, Haiyun Qian, Jian Wu, Bing-Lin Gu, and Wenhui Duan: 
Z. Li, H. Qian, J. Wu, B.-L. Gu, and W. Duan: 
\jo{\PRL}{100}{2008}{206802}.
%-----------
\bibitem{Cresti}
A. Cresti, G. Grosso, and G. P. Parravicini: 
\jo{\PRB}{77}{2008}{233402}. 
%----------
\bibitem{Nakabayashi}
J. Nakabayashi, D. Yamamoto, and S. Kurihara:
\jo{\PRL}{102}{2009}{066803}.
%----------
\bibitem{Ranis}
D. Rainis, F. Taddei, F. Dolcini, M. Polini, and R. Fazio:
\jo{\PRB}{79}{2009}{115131}. 
%----------
\bibitem{Mochizuki-1}
Y. Mochizuki and H. Yoshioka: 
\jo{\JPSJ}{78}{2009}{123701}. 
%----------
\bibitem{Mochizuki-2}
Y. Mochizuki and H. Yoshioka: 
\jo{\PHE}{42}{2010}{722}.
%--------------
\bibitem{Yoshioka-Higashibata}
H. Yoshioka and S. Higashibata:
\jo{J. Phys: Conf. Ser.}{150}{2009}{022105}. 
%----------









%------------
%\bibitem{Mochizuki}
%Y. Mochizuki: unpublished.
\end{thebibliography}
\end{document}